\documentclass[aps,pra,twocolumn,amsmath,amssymb]{revtex4}

\usepackage{graphicx}
\usepackage{dcolumn}
\usepackage{bm}

\begin{document}

\title{Manipulation and Detection of a Trapped Yb$^+$ Ion Hyperfine Qubit}

\author{S. Olmschenk$^{1,2}$}
	\email{smolms@umich.edu}
\author{K. C. Younge$^1$}
\author{D. L. Moehring$^1$}
\author{D. Matsukevich$^{1,2}$}
\author{P. Maunz$^{1,2}$}
\author{C. Monroe$^{1,2}$}
	\affiliation{${}^1$FOCUS Center and Department of Physics, University of Michigan, Ann Arbor, MI  48109 \\ ${}^2$JQI and Department of Physics, University of Maryland, College Park, MD 20742}
\date{\today}

\begin{abstract}
We demonstrate the use of trapped ytterbium ions as quantum bits for quantum information processing.  We implement fast, efficient state preparation and state detection of the first-order magnetic field-insensitive hyperfine levels of $^{171}$Yb$^+$, with a measured coherence time of 2.5 seconds.  The high efficiency and high fidelity of these operations is accomplished through the stabilization and frequency modulation of relevant laser sources.
\end{abstract}

\maketitle

\section{\label{sec:intro}Introduction}

Ions stored in radio-frequency (rf) traps have long been recognized as a promising implementation of quantum bits (qubits) for quantum information processing~\cite{cirac:cold-ions,NIST-JRes}. This is due in part to the long trapping lifetimes and long coherence times of particular internal electronic states of the ions. The ytterbium (Yb$^{+}$) ion is especially attractive because the strong ${}^2S_{1/2} \leftrightarrow {}^2P_{1/2}$ electronic transition near 369.53 nm is suitable for use with optical fibers, making schemes that require the coupling of atomic (hyperfine) qubits to photonic (optical) qubits feasible~\cite{duan:dlcz,simon:dist-entangle,boris:nature,duan:prob-photon,mckeever:atom-cavity,duan:freq-qubit,sherson:atom-light,beugnon:interference,volz:atom-photon,peter:interference,dave:ion-ion,jenkins:qtel,chou:func-ensemble,hijlkema:photon-server,wilk:atom-photon,wilk:pol-photon}.  Moreover, the large fine structure splitting of Yb$^{+}$ makes it amenable to schemes requiring fast manipulation with broadband laser pulses~\cite{poyatos:strong-excite,garcia-ripoll:fast-gates,duan:fast-gates,martin:ultrafast-rabi}.  Finally, the spin-1/2 nucleus of ${}^{171}$Yb$^{+}$ allows for simple, fast, and efficient preparation and detection of the ground state hyperfine levels.

\begin{figure}[!b]
\centering
\includegraphics[width=1.0\columnwidth,keepaspectratio]{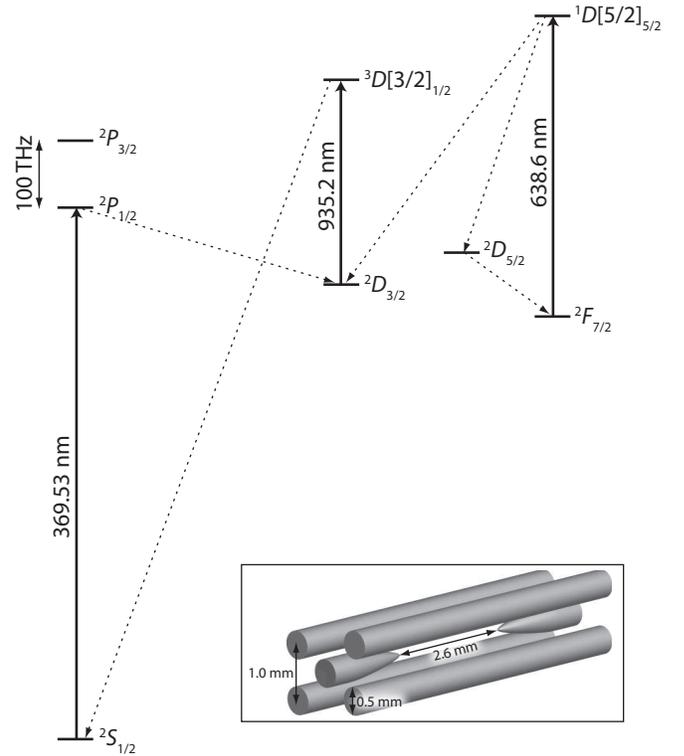}
\caption{Partial level scheme of Yb$^+$ (drawn to scale). The ${}^2S_{1/2} \leftrightarrow {}^2P_{1/2}$ transition at 369.53 nm is used for Doppler-cooling and to observe ion fluorescence.  The branching ratio from ${}^2P_{1/2}$ to the low-lying metastable ${}^2D_{3/2}$ state is measured to be $0.00501(15)$ (see App.~\ref{app:branching-ratio}).  This metastable state is depopulated by light at 935.2 nm resonant with the ${}^2D_{3/2} \leftrightarrow {}^3D[3/2]_{1/2}$ transition, which quickly returns the ion to the cooling transition~\cite{bell:four-level,cowan:spectra}.  The ion also reaches the ${}^2F_{7/2}$ state on the order of a few times per hour, at which point it is returned to the cooling cycle using light near 638.6 nm.  Inset:  schematic of the linear rf-Paul trap used in this experiment.}
\label{levelscheme}
\end{figure}

The ytterbium ion has previously been exploited by several groups in high-resolution spectroscopic studies~\cite{pendrill:hyperfine, fisk:171s12, roberts:171d52, tamm:171d32, yu:lifetime, blythe:171f, schneider:comp171} and for applications in quantum information~\cite{hannemann:self-learning, balzer:ybqip}.  In this paper, we report efficient preparation and measurement of a qubit stored in the first-order magnetic field-insensitive hyperfine levels of the ground state of $^{171}$Yb$^{+}$.  The high efficiency and high fidelity of these processes is achieved through the stabilization of the relevant lasers to an absorption line of molecular iodine near 739 nm, and frequency modulation of certain light sources.  We measure the coherence time of the qubit to be 2.5(3) seconds.  In addition, we provide improved measurements of the $^{2}D_{3/2}$ and $^{3}D[3/2]_{1/2}$ hyperfine splittings, and the branching ratio of spontaneous emission from $^2P_{1/2}$ to ${}^2D_{3/2}$.

\section{\label{sec:trap}Trapping Ytterbium Ions}

The ion trap employed in this experiment is a four-rod linear rf-Paul trap (Fig.~\ref{levelscheme}, inset).  The rods of the trap are 0.5 mm in diameter with a center to center spacing of 1.0 mm and end-cap spacing of 2.6 mm. The rf drive frequency of the trap is $\Omega_T/2\pi = 37$ MHz with center of mass trapping frequencies of $(\omega_x,\omega_y,\omega_z)/2\pi \approx (1,1,0.2)$ MHz.  Residual micromotion at the rf drive frequency is analyzed using a fluorescence cross-correlation technique~\cite{berkeland:micromotion}, and suppressed by applying static offset voltages to the trap rods~\cite{dehmelt:iontrap}.

Ytterbium (Yb$^{+}$) ions are loaded into the trap by photoionization of an atomic beam of neutral Yb.  A thermal beam of Yb atoms is produced by resistively heating a stainless steel tube filled with Yb metal, and is directed toward the trap.  A continuous-wave (cw) diode laser that provides approximately 5 mW of light at 398.91 nm is tuned to the ${}^1S_0 \leftrightarrow {}^1P_1$ transition of neutral Yb and is focused through the center of the trap with a waist of $\approx$50 $\mu$m.  A second beam near 369.53 nm is generated by frequency doubling the light produced by a cw amplified diode laser near 739.05 nm.  Approximately 375 mW of 739.05 nm light is sent to a resonant cavity containing a critically phase-matched LBO crystal, producing more than 14 mW at 369.53 nm.  The majority of this light at 369.53 nm is aligned counter-propagating to the beam at 398.91 nm, and is focused through the trap with a waist of $\approx$30 $\mu$m.  Neutral Yb atoms passing through these beams are photoionized by way of a resonantly assisted, dichroic, two-photon transition \cite{balzer:ybqip}: the 398.91 nm light excites Yb atoms from the ${}^1S_0$ to the ${}^1P_1$ level, from which the 369.53 nm light can promote the electron to the continuum.  The neutral Yb beam is approximately perpendicular to the 398.91 nm beam to minimize Doppler shifts and allow for isotopically selective loading.

After photoionization of a neutral atom in the trap, the confined Yb$^{+}$ ion is Doppler-cooled by the light at 369.53 nm, which is slightly red-detuned of the ${}^2S_{1/2} \leftrightarrow {}^2P_{1/2}$ transition, as depicted in Fig.~\ref{levelscheme}.  The ${}^2P_{1/2}$ state also decays to the metastable ${}^2D_{3/2}$ level, and we measure this branching ratio to be $0.00501(15)$ (see App.~\ref{app:branching-ratio} for details).  In order to maintain fluorescence and cooling, light at 935.2 nm is used to drive the atom from the ${}^2D_{3/2}$ to the ${}^3D[3/2]_{1/2}$ level, from which it quickly returns to the ${}^2S_{1/2}$ ground state~\cite{bell:four-level}.  An additional complication arises from the presence of the low-lying ${}^2F_{7/2}$ state.  There are no allowed decays from the four levels used in cooling to ${}^2F_{7/2}$.  Even so, the ion falls into this state a few times per hour, probably due to collisions with residual background gas~\cite{lehmitz:pop-trap,bauch:pop-trap}.  Laser light near 638.6 nm depopulates the ${}^2F_{7/2}$ level, returning the ion to the four-level cooling cycle.  Doppler-cooling localizes the ion to within the diffraction-limited resolution of our imaging optics, but not to within the Lamb-Dicke limit.  The mean storage time of a trapped ion is several days.

Photons at 369.53 nm scattered by the ion are collected by an objective lens with a numerical aperture of 0.27.  An aperture at the intermediate image generated by the objective reduces observed background light.  This image is re-imaged using a doublet lens and directed to either a photon-counting photomultiplier tube (PMT) or camera.  The image provided by the camera is used to monitor the loading process and subsequently confirm the presence of a single Yb$^{+}$ ion in the trap.  The PMT, with its higher detector efficiency, is used to measure the ion fluorescence during state detection.

\section {\label{sec:lasers}Laser Stabilization}

\begin{figure}
\centering
\includegraphics[width=1.0\columnwidth,keepaspectratio]{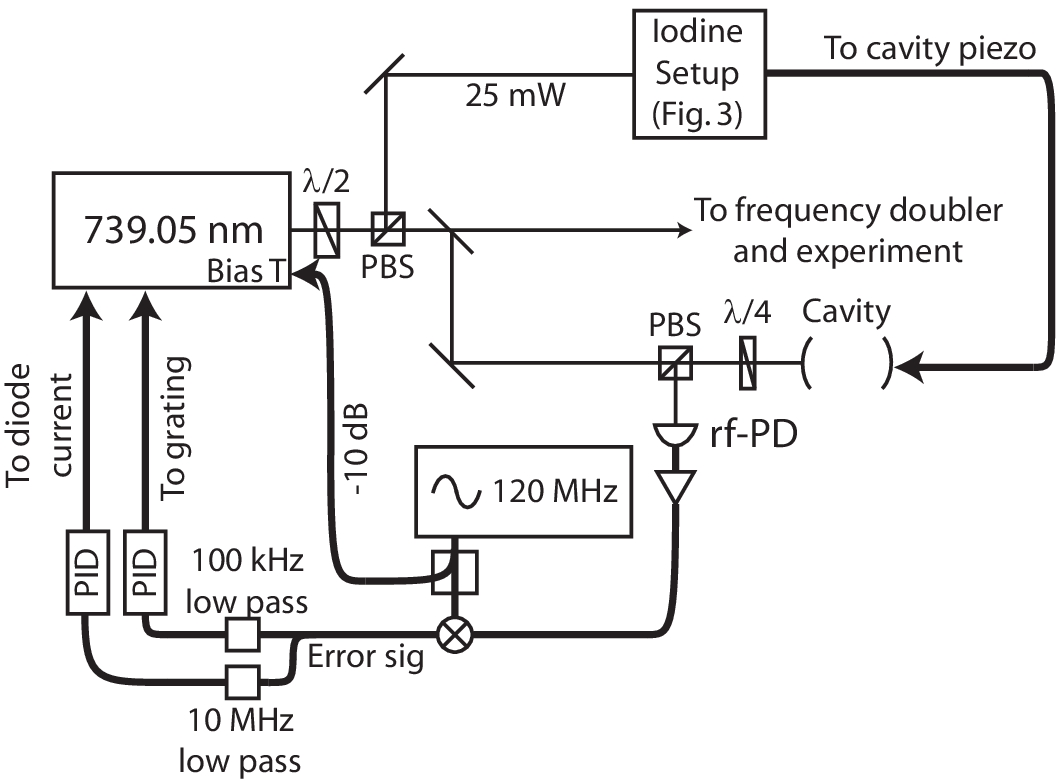}
\caption{Experimental apparatus for locking the 739.05 nm laser to an iodine-stabilized cavity.  The current of the diode laser is weakly modulated at 120 MHz via a Bias T.  Approximately 25 mW of 739.05 nm light is diverted from the main 400 mW beam to the iodine setup (Figure~\ref{iodine}).  The error signal generated by the iodine setup is sent to a proportional-integral-derivative (PID) servo controller, the output of which is sent to the piezo of the reference cavity.  An additional 120 $\mu$W of the main 739.05 nm beam is sent to this reference cavity. The cavity reflection is measured with a radio-frequency photodiode (rf-PD), and the resulting electronic signal is sent through an amplifier and mixed with a 120 MHz signal to produce an error signal.  The lock is achieved by feeding back to both the laser grating and diode current with appropriate filter electronics.  In the figure: PBS is a polarizing beamsplitter; $\lambda/2$ is a half waveplate; and $\lambda/4$ is a quarter waveplate.}
\label{739lock}
\end{figure}

Efficient quantum operations on Yb$^+$ ions require that the 739.05 nm and 935.2 nm lasers are stable in frequency to well within the linewidths of the respective transitions in the experiment.  To achieve this, the 739.05 nm laser is locked to a passively stable reference cavity using an rf (Pound-Drever-Hall) lock \cite{drever:stab,black:pdh}.  We then use saturated-absorption spectrocopy of iodine to stabilize the length of the cavity to an absolute frequency reference.  Lastly, we lock the 935.2 nm laser to the same cavity using a side-of-fringe technique.

The relevant optics and electronics for the rf stabilization of the 739.05 nm laser to the cavity are shown in Fig.~\ref{739lock}.  The current of the diode laser is weakly modulated at 120 MHz, and approximately 120 $\mu$W of the main 739.05 nm beam is sent to the reference cavity~\cite{doubling-note}.  The cavity reflection is measured with a radio-frequency photodiode, and the signal is then amplified and mixed with the 120 MHz reference oscillator.  The resulting error signal is used to stabilize the 739.05 nm laser to the reference cavity by feedback control of the laser diode current and grating angle.  The cavity is passively stable on short time-scales, and for long-term stability it is locked to an absorption line of molecular iodine.

\begin{figure}
\centering
\includegraphics[width=1.0\columnwidth,keepaspectratio]{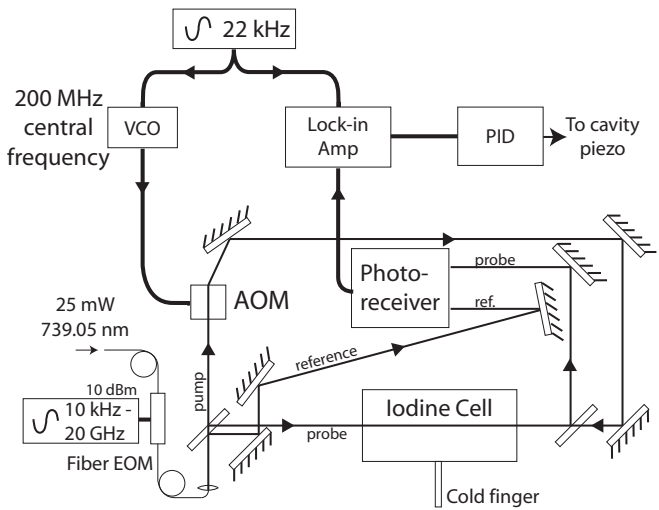}
\caption{Setup for locking to molecular iodine via saturated absorption spectroscopy. About 25 mW of 739.05 nm light is incident on the fiber-coupled electro-optic modulator (EOM) with 6 mW of this light transmitted.  Approximately 1/3 of the power is transferred into each of the two first-order sidebands.  The beam is then split into reference, probe, and pump beams by a glass plate.  The pump beam passes through an acousto-optic modulator (AOM) (driven by a voltage-controlled oscillator (VCO)) which is frequency modulated at 22 kHz, imprinting this modulation on the pump beam.  The pump and probe beams are counter-propagating and focused through the iodine vapor cell.  Ultimately, the probe and reference beams are incident on a photoreceiver.  A lock-in amplifier mixes the 22 kHz signal with the signal from the photoreceiver, and the resulting error signal is sent to the proportional-integral-derivative (PID) controller that controls the cavity length.}
\label{iodine}
\end{figure}

\begin{figure}
\centering
\includegraphics[width=1.0\columnwidth,keepaspectratio]{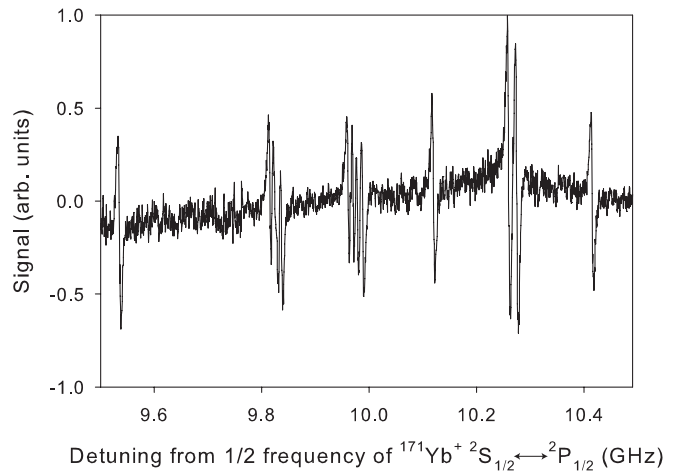}
\caption{Hyperfine structure of the iodine absorption line approximately 10 GHz~\cite{our-i2-lines-note} detuned from half the frequency of the ${}^2S_{1/2} \leftrightarrow {}^2P_{1/2}$ transition of $^{171}$Yb$^{+}$ (739.05 nm).  Each point is separated in frequency by 0.5 MHz, and was integrated for about 1.5 seconds.  The absolute positions of the features are accurate to within 20 MHz, while the frequency splittings between iodine resonances are accurate to within 5 MHz.}
\label{hyperfine}
\end{figure}

Molecular iodine is often chosen as a frequency reference for wavelengths from the near-infrared (e.g. 830 nm~\cite{ludvigsen:830i2}) to the disassociation limit at 499.5 nm because of the density of narrow absorption lines in this region~\cite{i2-atlas2, i2-atlas1}.  These lines can serve as excellent frequency references for laser stabilization to a few parts in $10^{-9}$ or better~\cite{edwards:iodine}.  However, in the region applicable to Yb$^+$ (739.05 nm) most of the lines are weak at room temperature.  In order to thermally achieve sufficient population of the rovibrational levels of the molecule that are the lower states of these ($B \rightarrow X$) transitions, the iodine must be heated to over 600 K~\cite{dube:francium}.  In our setup, we employ a homemade iodine vapor cell 20 cm in length with two quartz windows 25 mm in diameter and a 5 cm long cold finger.  The cold finger is isolated from the heating elements such that it remains at room temperature, which is necessary to avoid pressure broadening of the transition~\cite{dube:francium}.

We stabilize the reference cavity to molecular iodine using standard saturated absorption spectroscopy, as depicted in Figure~\ref{iodine}.  The 739.05 nm diode laser produces about 400 mW of output power, 25 mW of which is diverted to the iodine setup.  This light is sent to a 20 GHz bandwidth fiber electro-optic modulator (EOM)~\cite{eom-note}, and one of the resulting first-order sidebands is used for iodine spectroscopy.  At 739.05 nm, 25 mW of light can be safely sent through the EOM without photorefractive damage to the crystal.  However, due to coupling and transmission losses, only 6 mW of this light is transmitted.  The modulator receives approximately 10 dBm of rf power, which transfers about 1/3 of the power (effectively, $\approx$2 mW) into the first-order sideband used for spectroscopy~\cite{sideband-note}.  A feature of this setup is that the large bandwidth of the fiber EOM allows the laser to be scanned over a wide range while remaining locked to a given iodine absorption line.  

In order to determine the best absorptive feature for use as an absolute frequency reference, we performed spectroscopy on three iodine reference lines.  Absorption lines of molecular iodine are observed at detunings of approximately 13 GHz, 10 GHz, and -5 GHz~\cite{our-i2-lines-note} from the target wavelength of 739.05 nm (twice the wavelength of the ${^2S_{1/2}} \leftrightarrow {^2P_{1/2}}$ transition of $^{171}$Yb$^+$).  The continuous tuning of the fiber EOM over nearly 20 GHz enables the spectroscopic measurement of the hyperfine structure of these absorption lines.  For this measurement, we stabilize the 739.05 nm laser to the fluorescence of a single $^{174}$Yb$^+$ ion using a side-of-fringe technique.  We estimate that this results in a stability of better than 5 MHz over the course of the measurement.  As the intensity of the 935.2 nm light on the ion was far above saturation ($\sim$$20 \times I_{sat}$), small fluctuations in the frequency of the 935.2 nm laser had a negligible effect on the ion fluorescence.

With the 739 nm laser locked to a trapped ion, the rf applied to the fiber EOM was varied in 0.5 MHz steps over the areas of interest, and the output of the lock-in amplifier was recorded.  Because of the weak nature of the lines being investigated, each point was integrated for about 1.5 seconds using a lock-in time constant of 300 ms to achieve a reasonable signal-to-noise ratio.  We mapped the hyperfine structure of the three aforementioned absorption lines, with the hyperfine structure of the line near a detuning of 10 GHz (739.03 nm) presented in Fig.~\ref{hyperfine}.  The doublet feature at a detuning of approximately 10.3 GHz produces a strong signal, and can be used to stabilize the reference cavity of Fig.~\ref{739lock}.

After the cavity has been stabilized to iodine, a weak ($\approx$500 $\mu$W) beam at 935.2 nm is sent through a fiber EOM and then directed ($\approx$50 $\mu$W) to the cavity.  The rf used to drive this fiber EOM is tuned such that one of the resulting first-order sidebands is resonant with the cavity.  The transmitted light is observed with a photodiode, and the consequent signal sent to a PID controller.  In this way, we lock the 935.2 nm laser to the iodine-stabilized cavity using a side-of-fringe technique.

\section{\label{sec:qubit}The Ytterbium Qubit}

The requirements for the physical implementation of quantum computation include state initialization and state detection of the quantum bits (qubits), a set of universal gates, and long qubit coherence time~\cite{divincenzo}.  We demonstrate that the ${}^2S_{1/2}$ hyperfine levels of $^{171}$Yb$^+$ may serve as an excellent qubit for quantum information processing.  By applying additional laser frequency sources according to the hyperfine splittings of the relevant energy levels, we are able to achieve fast, efficient state initialization and state detection, reliable single-qubit operations, and a qubit coherence time that far exceeds the duration of these processes.

\begin{figure}
\centering
\includegraphics[width=1.0\columnwidth,keepaspectratio]{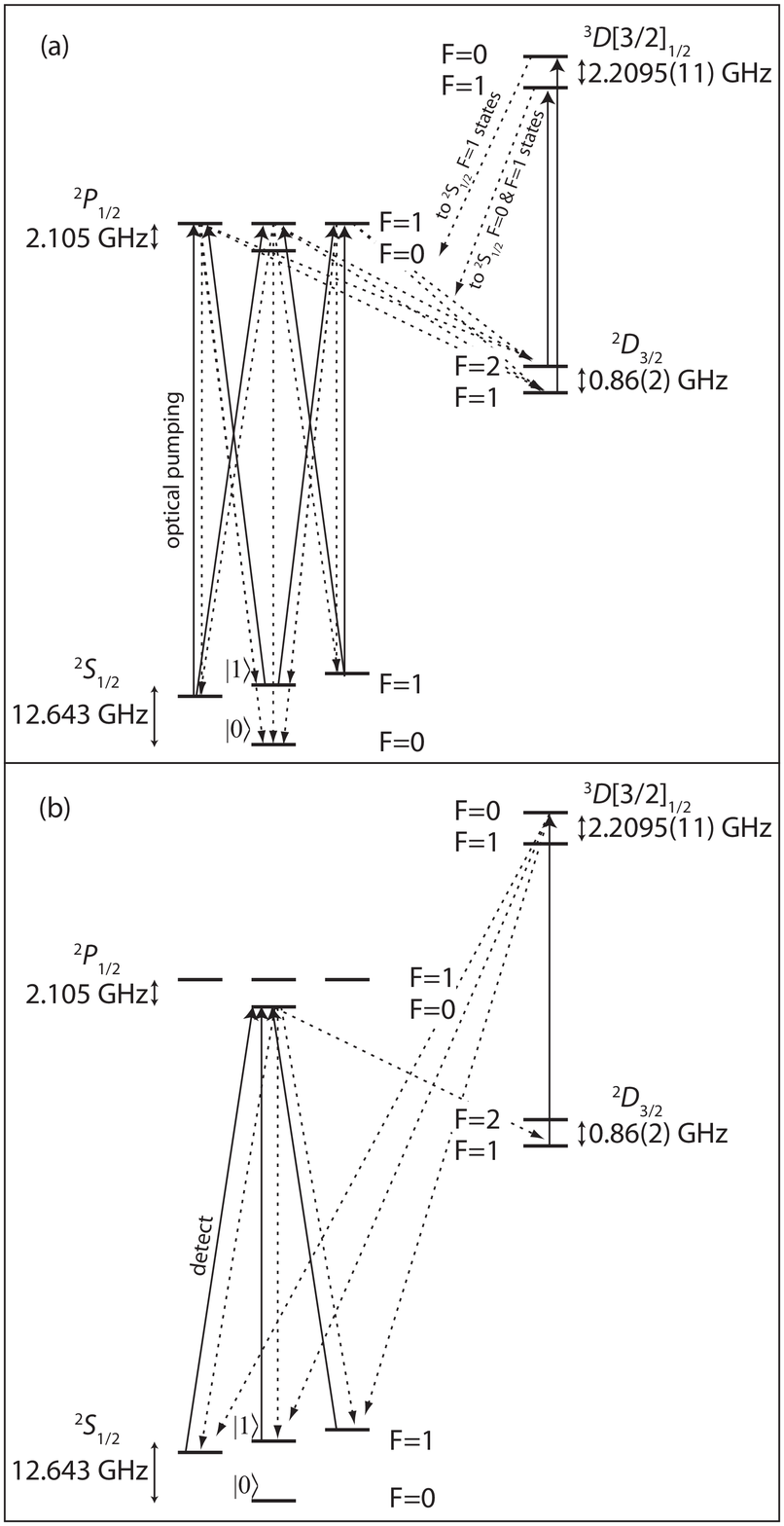}
\caption{The $^{171}$Yb$^+$ qubit (not to scale). The ${}^2S_{1/2} \vert F=1,m_F=0 \rangle$ state is defined to be $\vert 1 \rangle$, and the ${}^2S_{1/2} \vert F=0,m_F=0 \rangle$ state is defined to be $\vert 0 \rangle$.  (a) State initialization to $\vert 0 \rangle$ by application of light resonant with the ${}^2S_{1/2} \vert F=1 \rangle \leftrightarrow {}^2P_{1/2} \vert F=1 \rangle$ transition.  (b) Detection of the qubit state.  If the qubit state is $\vert 1 \rangle$, the 369.53 nm light applied for detection is nearly on resonance, and the ion scatters many photons.  If the state is $\vert 0 \rangle$, very few photons are scattered.  Measurements of the hyperfine splitting of the ${}^2S_{1/2}$ and ${}^2P_{1/2}$ levels are found in Ref.~\cite{fisk:171s12} and~\cite{pendrill:hyperfine}, respectively.  Details regarding our measurement of the hyperfine splitting of ${}^2D_{3/2}$ and ${}^3D[3/2]_{1/2}$ are found in App.~\ref{app:hyperfine}.}
\label{qubit}
\end{figure}

The relevant energy levels for the $^{171}$Yb$^+$ ion are shown in Fig.~\ref{qubit}.  The qubit is defined to be the two first-order magnetic field-insensitive hyperfine levels of the ${}^2S_{1/2}$ ground state.  We define the ${}^2S_{1/2} \vert F=1,m_F=0 \rangle$ state as the logical qubit state $\vert 1 \rangle$, and the ${}^2S_{1/2} \vert F=0,m_F=0 \rangle$ state as $\vert 0 \rangle$.  Here $F$ is the total angular momentum of the atom and $m_F$ is its projection along the quantization axis.  The qubit states are separated by a frequency of 12,642,812,118.5 + $\delta$ Hz, where $\delta = (310.8) B^2$ is the second-order Zeeman shift, and $B$ is the magnetic field in gauss~\cite{fisk:171s12}.

\begin{figure}
\centering
\includegraphics[width=1.0\columnwidth,keepaspectratio]{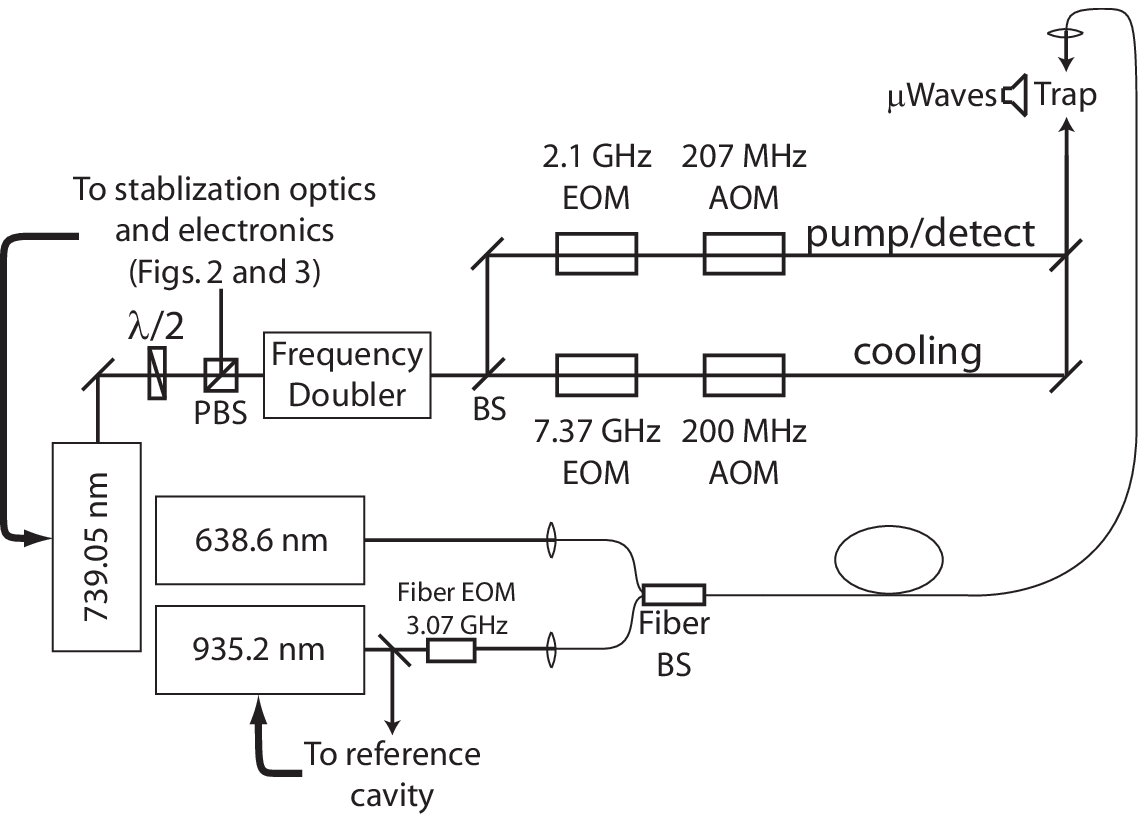}
\caption{The experimental setup for initialization, manipulation, and detection of the ${}^{171}$Yb$^{+}$ qubit.  The 7.37 GHz EOM in the cooling beam is tuned so that the positive second-order sideband is on resonance with the ${}^2S_{1/2} \vert {F=0} \rangle \leftrightarrow {}^2P_{1/2} \vert {F=1} \rangle$ transition, preventing population trapping in $\vert 0 \rangle$ during cooling.  The EOM in the pump/detect beam at 2.1 GHz is used to optically pump the ion into $\vert 0 \rangle$ during state initialization.  The 935.2 nm beam passes through a widely tunable fiber EOM, which is used to generate a frequency component to depopulate the ${}^2D_{3/2} \vert {F=2} \rangle$ manifold during cooling and optical pumping.  The AOMs are used to quickly turn the beams on and off.  The AOM in the cooling beam is tuned such that the diffracted beam is about half a linewidth ($\approx$10 MHz) away from resonance to achieve efficient cooling.  The pump/detect beam is tuned only $\approx$3 MHz from resonance to generate more scattering events, reducing the time needed to distinguish between qubit states with high fidelity.  Microwaves ($\mu$Waves) at about 12.643 GHz are sent to the trap for single-qubit manipulation.}
\label{exp_setup_1trap}
\end{figure}

The ${}^{171}$Yb$^{+}$ ion is Doppler-cooled by light slightly red-detuned ($\approx$10 MHz) of the ${}^2S_{1/2} \vert F=1 \rangle \leftrightarrow {}^2P_{1/2} \vert F=0 \rangle$ transition at 369.53 nm.  During cooling, there is approximately 6 $\mu$W of 369.53 nm light focused to a waist of $\approx$30 $\mu$m at the center of the trap.  Off-resonant coupling to the ${}^2P_{1/2} \vert F=1 \rangle$ manifold results in population trapping in $\vert 0 \rangle$.  To prevent this during cooling intervals, the 369.53 nm cooling beam is passed through a bulk resonant EOM driven at 7.37 GHz, as shown in Fig.~\ref{exp_setup_1trap}.  The resulting positive second-order sideband is resonant with the ${}^2S_{1/2} \vert F=0 \rangle \leftrightarrow {}^2P_{1/2} \vert F=1 \rangle$ transition, which returns the ion to the cooling cycle~\cite{crap-eom-note}.  Population trapping in the ${}^2D_{3/2} \vert F=1 \rangle$ manifold is avoided by application of a laser at 935.2 nm, which rapidly returns the atom to the cooling cycle via the ${}^3D[3/2]_{1/2} \vert F=0 \rangle$ level~\cite{bell:four-level}.  About 5 mW of 935.2 nm light is sent to the trap, focused to a spot with waist $\approx$200 $\mu$m~\cite{power-note}.  When the ion is excited to the ${}^2P_{1/2} \vert F=1 \rangle$ manifold, which occurs off-resonantly during cooling, the ion may decay to ${}^2D_{3/2} \vert F=2 \rangle$.  To depopulate this level, we use a fiber EOM driven at 3.07 GHz to add a frequency component to the 935.2 nm light that is resonant with the ${}^2D_{3/2} \vert F=2 \rangle \leftrightarrow {}^3D[3/2]_{1/2} \vert F=1 \rangle$ transition (see App.~\ref{app:hyperfine}).  Finally, the ${}^2F_{7/2}$ hyperfine levels are depopulated by laser light that is switched between the ${}^2F_{7/2} \vert F=3 \rangle \leftrightarrow {}^1D[5/2]_{5/2} \vert F=2 \rangle$ and ${}^2F_{7/2} \vert F=4 \rangle \leftrightarrow {}^1D[5/2]_{5/2} \vert F=3 \rangle$ transitions, both near 638.6 nm~\cite{roberts:171d52}.  Nearly 2 mW of 638.6 nm light is present at the trap, focused to $\approx$200 $\mu$m.  The ion resides in a magnetic field of about 5 gauss for definition of the quantization axis and to avoid coherent population trapping~\cite{berkeland:coherent-pop}.

State initialization is accomplished by optically pumping the ${}^{171}$Yb$^{+}$ ion into the $\vert 0 \rangle$ state, as shown in Fig.~\ref{qubit}(a).  A beam at 369.53 nm is passed through a bulk resonant 2.1 GHz EOM before reaching the ion.  During state initialization, this 2.1 GHz EOM is switched on~\cite{pump-note}, generating a positive first-order sideband resonant with the ${}^2S_{1/2} \vert F=1 \rangle \leftrightarrow {}^2P_{1/2} \vert F=1 \rangle$ transition.  From the ${}^2P_{1/2} \vert F=1 \rangle$ manifold, the ion has a 1/3 chance of decaying to $\vert 0 \rangle$.  Since ${}^2P_{1/2} \vert F=1 \rangle$ also decays to ${}^2D_{3/2} \vert F=2 \rangle$, the aforementioned 3.07 GHz sideband in the 935.2 nm light is essential for efficient state initialization.  With approximately 50 $\mu$W of 369.53 nm light focused to $\approx$30 $\mu$m at the trap, the ion is optically pumped to the $\vert 0 \rangle$ state in less than 500 ns with near perfect efficiency.

\begin{figure}
\centering
\includegraphics[width=1.0\columnwidth,keepaspectratio]{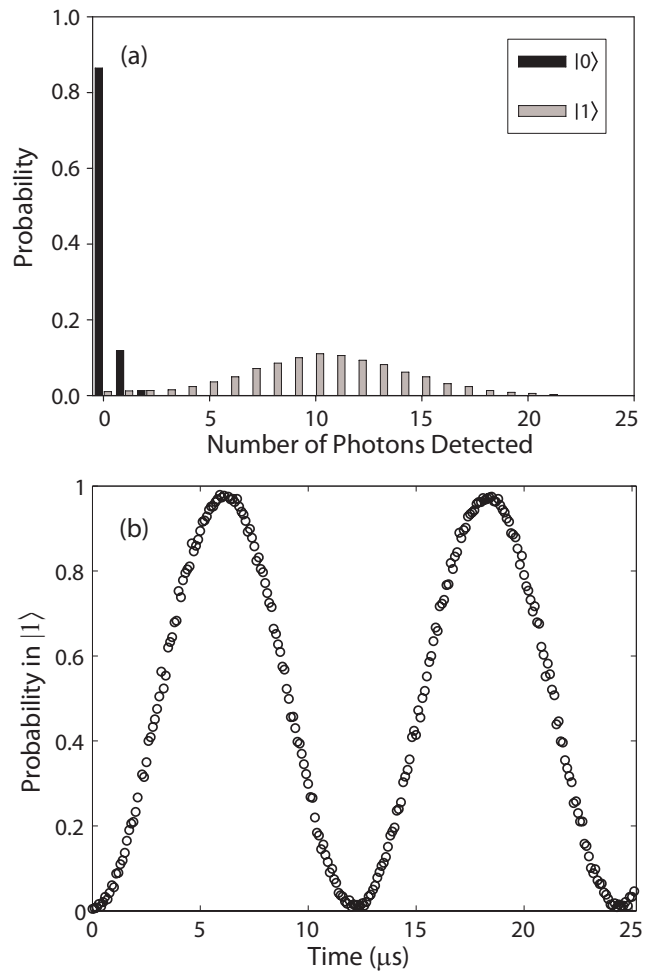}
\caption{(a) Histograms of detected photons after the ion is prepared in each of the two qubit states.  The detection interval is 1000 $\mu$s, with 0.8 $\mu$W of 369.53 nm light focused to $\approx$30 $\mu$m.  \textbf{(Black histogram)} The ion is initialized to the $\vert 0 \rangle$ state.  Transitions to ${}^2P_{1/2} \vert F=0 \rangle$ are forbidden by selection rules, the incident light is detuned from ${}^2P_{1/2} \vert F=1 \rangle$ by the combined hyperfine splitting of 14.7 GHz, and therefore the atom scatters very few photons.  The histogram is the result of 15290 measurements.  \textbf{(Grey histogram)} The ion is initialized to $\vert 0 \rangle$, and then rotated to $\vert 1 \rangle$ by application of microwaves.  During the detection interval, the incident light is on resonance with the ${}^2S_{1/2} \vert F=1 \rangle \rightarrow {}^2P_{1/2} \vert F=0 \rangle$ transition and the atom scatters many photons.  The histogram is the result of 16497 measurements.  (b) Rabi oscillations between the $\vert 0 \rangle$ and $\vert 1 \rangle$ states by application of microwaves at 12.642821 GHz. The time required to drive a $\pi$-pulse is 6.0 $\mu$s.  Each point is the result of 1000 measurements.}
\label{dark-vs-bright-hist}
\end{figure}

Accurate detection of the qubit state is a critical step in quantum computation and communication protocols.  In ${}^{171}$Yb$^{+}$, state detection is accomplished using standard ion fluorescence techniques, as illustrated in Fig.~\ref{qubit}(b)~\cite{NIST-JRes,boris:hyperfine,mark:detection}.  The 369.53 nm light of the detect beam is tuned to be nearly on resonance with the ${}^2S_{1/2} \vert F=1 \rangle \leftrightarrow {}^2P_{1/2} \vert F=0 \rangle$ transition.  If the ion is prepared in the state $\vert 0 \rangle$, this incident light is detuned from ${}^2P_{1/2} \vert F=1 \rangle$ by 14.7 GHz, and thus the ion scatters very few photons.  Conversely, if the $\vert 1 \rangle$ state is prepared, then the impinging light is nearly on resonance, and many scattered photons are observed (see Fig.~\ref{dark-vs-bright-hist}a).  The state of the ion is determined by the number of photons observed by the PMT during the detection interval.  In the experiments presented here, if more than one photon is observed during detection, the ion is defined to be in the state $\vert 1 \rangle$; if one or zero photons are detected, the ion is defined to be in $\vert 0 \rangle$.

The limiting source of error in state detection is off-resonant coupling to the ${}^2P_{1/2} \vert F=1 \rangle$ manifold.  A detailed study of the theoretical detection fidelity limit for different ion species is presented in Ref.~\cite{mark:detection}.  Given a photon collection efficiency of 0.001 (one out of every 1000 scattered photons is collected and detected) in the limit of low intensity of the detect beam, the maximum detection fidelity is calculated to be $99.51\%$ for $^{171}$Yb$^+$~\cite{detect-note}.  The majority of the error is a consequence of off-resonant excitation to the ${}^2P_{1/2} \vert F=1 \rangle$ manifold when the ion has been prepared in $\vert 1 \rangle$, as the hyperfine splitting of the ${}^2P_{1/2}$ level is only 2.1 GHz.  Instead, if the initial state is $\vert 0 \rangle$, the incident light is detuned from the ${}^2P_{1/2} \vert F=1 \rangle$ manifold by 14.7 GHz and transitions to the ${}^2P_{1/2} \vert F=0 \rangle$ level are forbidden by selection rules.  Inclusion of decay into the ${}^2D_{3/2}$ level in this calculation reduces the total number of 369.53 nm photons scattered by the ion.  In addition, off-resonant coupling to the ${}^3D[3/2]_{1/2} \vert F=1 \rangle$ level may occur while depopulating the ${}^2D_{3/2} \vert F=1 \rangle$ manifold.  However, given the small branching ratio into ${}^2D_{3/2}$, we estimate these additions to the calculation constitute a negligible change in the theoretical detection fidelity ($<0.01\%$).  A far more important issue is the reduced scattering rate of the $^{171}$Yb$^+$ ion due to coherent population trapping in the ${}^2S_{1/2} \vert F=1 \rangle$ manifold~\cite{berkeland:coherent-pop}.  In optimal conditions, the scattering rate of the ${}^2S_{1/2} \vert F=1 \rangle \leftrightarrow {}^2P_{1/2} \vert F=0 \rangle$ transition used for detection is reduced to about 1/3 the natural rate, while the error rate remains virtually unchanged.  Hence, for a photon collection efficiency of 0.001, the theoretical detection fidelity is reduced to about $98.55\%$ (Table~\ref{tab:det-fidelity}).

\begin{table}
	\caption{\label{tab:det-fidelity}Theoretical state detection fidelity in ${}^{171}$Yb$^+$ for various photon collection efficiencies when the reduced scattering rate due to coherent dark states is included.}
	\begin{ruledtabular}
		\begin{tabular}{cc}
			{} & Theoretical \\
			Photon collection efficiency & state detection fidelity \\
			\hline
			0.001 & 98.55\% \\
			0.003 & 99.51\% \\
			0.01 & 99.85\% \\
			0.03 & 99.95\% \\
			0.1 & 99.985\% \\
		\end{tabular}
	\end{ruledtabular}
\end{table}

We experimentally optimize state detection with respect to the intensity and duration of the light applied during the detect interval.  As a result of the small branching ratio into ${}^2D_{3/2}$, we observe that the fidelity of state detection has negligible dependence on the 935.2 nm intensity (above 500 $\mu$W, focused to $\approx$200 $\mu$m).  The 369.53 nm light was varied over a wide range of intensities and durations.  Optimum state detection was realized with 0.8 $\mu$W of 369.53 nm light, focused to a spot of radius $\approx$30 $\mu$m at the trap, and incident on the ion for 1000 $\mu$s.  The resulting histograms for state preparation in $\vert 0 \rangle$ versus $\vert 1 \rangle$ are presented in Fig.~\ref{dark-vs-bright-hist}a.  Given these parameters, the state detection fidelity is measured to be 97.9(2)\%.

The length of the detection interval is important in this measurement, as dark counts of the PMT become significant (measured at $\approx$150 per second).  Consequently, the majority of the ``photons'' detected in the black histogram of Fig.~\ref{dark-vs-bright-hist}a are probably due to dark counts of the PMT.  Amending this technical issue may increase the measured state detection fidelity in future experiments.

After the ion is initialized to $\vert 0 \rangle$, we can rotate the state of the ion between $\vert 0 \rangle$ and $\vert 1 \rangle$ by applying microwaves resonant with the 12.643 GHz ${}^2S_{1/2}$ hyperfine splitting.  To accomplish this, several watts of microwave power is sent to a truncated waveguide that is 3 cm in length, located 7 cm from the center of the trap.  The resulting Rabi oscillations between $\vert 0 \rangle$ and $\vert 1 \rangle$ demonstrate single-qubit operations on the ${}^{171}$Yb$^{+}$ ion (Fig.~\ref{dark-vs-bright-hist}b).

\begin{figure}
\centering
\includegraphics[width=1.0\columnwidth,keepaspectratio]{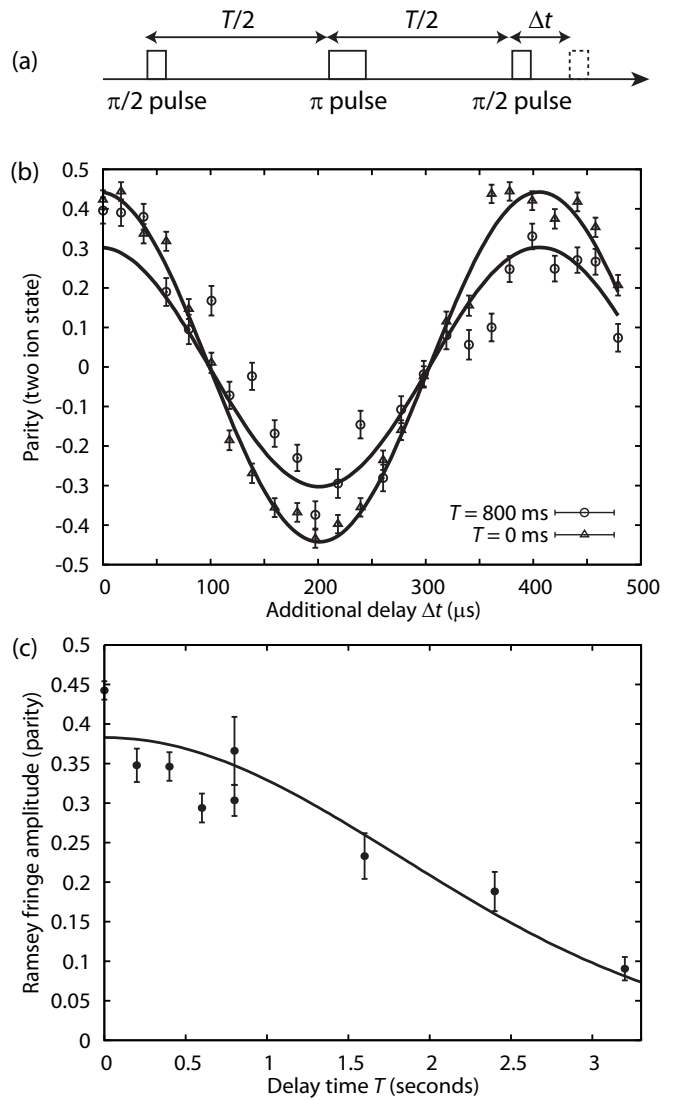}
\caption{Measurement of the coherence time of the ${}^{171}$Yb$^+$ qubit.  (a) The series of microwave pulses used for this measurement.  The two $\pi/2$ pulses are separated by a variable time $T$, to measure the decay in the amplitude of the Ramsey fringe as a function of time.  (b) Observed Ramsey fringes in the parity of the states of the two ions for delay times $T = 0$ ms and $T=800$ ms.  (c) Decay in the amplitude of the observed Ramsey fringe as function of delay time $T$.  The line is a Gaussian fit to the data, with a 1/e decay time of 2.5(3) seconds.}
\label{fig:coherence}
\end{figure}

Finally, we perform a Ramsey-type experiment to measure the coherence time of the qubit states.  In previous atomic clock work~\cite{fisk:171s12}, a coherence time exceeding 15 minutes has been achieved with these same qubit levels.  Since the coherence time of the qubit exceeds the stability of our microwave oscillator, we are unable to measure the coherence time of a single ion directly.  Instead, we measure the coherence of one ion with respect to a second trapped ion.  The two $^{171}$Yb$^{+}$ ions are stored in two independent (nearly identical) traps separated by about one meter.  A deliberate disparity in the magnetic field between the positions of the two ions alters the microwave transition frequency, which is 12.642821 GHz at one ion position, by about 2.43 kHz between the two ions.  To measure the qubit coherence time, both ions are first initialized to $\vert 0 \rangle$.  We then apply a microwave $\pi/2$ pulse to the two ions, wait for a time $T/2$, apply a microwave $\pi$ echo pulse to both ions, wait for another time $T/2+\Delta t$, apply a microwave $\pi/2$ analyzing pulse to both ions, and finally detect the state of each ion (see Fig.~\ref{fig:coherence}a).  The phase of the second microwave $\pi/2$ pulse is scrambled in order to eliminate the effect of the stability of the microwave oscillator.  As a consequence, no Ramsey fringe is present in the signal from a single ion.  However, a plot of the parity (the correlation between the measured states of the two ions) as a function of the additional delay $\Delta t$ displays a Ramsey fringe, as shown in Fig.~\ref{fig:coherence}b.  Lacking entanglement between the two ions, the amplitude of this fringe cannot exceed 0.5.  As the delay time $T$ is increased, the amplitude of the Ramsey fringe decreases (Fig.~\ref{fig:coherence}c).  Assuming gaussian fluctuations of the transition frequencies of the two ions, we fit the amplitude to a gaussian decay and obtain a 1/e coherence time of 2.5(3) seconds.  The measured coherence time is likely limited by fluctuations of the differential magnetic field between the two ion positions through the second-order Zeeman shift.  The coherence time should be significantly longer when lower static magnetic fields are used.

\section{Conclusion}

We have realized an experimental setup for quantum information processing with single trapped Yb$^{+}$ ions.  The relevant lasers are stabilized to an absorption line in molecular iodine using saturated absorption spectroscopy.  The branching ratio from ${}^2P_{1/2}$ to ${}^2D_{3/2}$ has been measured to be $0.00501(15)$.  In addition, the hyperfine splittings of the ${}^2D_{3/2}$ and ${}^3D[3/2]_{1/2}$ levels have been measured to be $0.86(2)$ GHz and $2.2095(11)$ GHz, respectively.  We have experimentally demonstrated fast, efficient state initialization, state detection fidelity of 97.9(2)\%, reliable single-qubit operations, and a qubit coherence time of 2.5(3) seconds.  Overall, the ${}^{171}$Yb$^{+}$ ion has been shown to be an excellent qubit candidate for applications in quantum information and quantum communication.

\begin{acknowledgments}
The authors would like to acknowledge useful discussions with M. Acton, P. C. Haljan, P. Blythe, and Chr. Wunderlich.  This work is supported by the National Security Agency and the Disruptive Technology Organization under Army Research Office contract W911NF-04-1-0234, and the National Science Foundation Information Technology Research and Physics at the Information Frontier Programs.
\end{acknowledgments}

\appendix
\section{\label{app:branching-ratio}${}^2P_{1/2}$ Branching Ratio}
We measure the ${}^2P_{1/2}$ branching ratio to the ${}^2D_{3/2}$ level by observing the decay in fluorescence of a single trapped ${}^{174}$Yb$^{+}$ ion.  For this experiment, the 739.05 nm laser is stabilized to an absorption feature in iodine via a reference cavity, as discussed in Sec.~\ref{sec:lasers}.  The 935.2 nm laser is stabilized to the same cavity, but here is tuned far ($\approx$3 GHz) from the ${}^2D_{3/2} \leftrightarrow {}^3D[3/2]_{1/2}$ transition.  On the way to the ion, the 935.2 nm light is passed through a fiber EOM that produces a first-order sideband resonant with the ${}^2D_{3/2} \leftrightarrow {}^3D[3/2]_{1/2}$ transition of ${}^{174}$Yb$^{+}$.  

The measurement sequence begins with a 100 $\mu$s interval during which the fiber EOM modulating the 935.2 nm light is on, and both the 935.2 nm light and 369.53 nm light are incident on the ion.  In a subsequent interval of 95 $\mu$s, the fiber EOM modulating the 935.2 nm light is switched off, thus ceasing depopulation of ${}^2D_{3/2}$.  During this time, photons scattered by the ion are collected and sent to the PMT, with the arrival times of the photons recorded by a time-to-digital converter.  The fluorescence signal has an exponential decay arising from population trapping in the ${}^2D_{3/2}$ state, with functional form $\exp{(-\mathcal{P}_{P_{1/2}} \gamma R t)}$.  Here, $\gamma$ is the linewidth of the ${}^2P_{1/2}$ state; $R$ is the branching ratio (probability of decay) into ${}^2D_{3/2}$; and $\mathcal{P}_{P_{1/2}}$ is the population of the ${}^2P_{1/2}$ state, which is a function of the power ($p$) of the 369.53 nm light at the ion (for a constant beam waist), given by $\mathcal{P}_{P_{1/2}} = (p/p_{sat})/(2(1+p/p_{sat}))$ for resonant light.  Decay from the metastable ${}^2D_{3/2}$ state is neglected, since the lifetime of this state (52.7 ms~\cite{yu:lifetime}) is orders of magnitude longer than the measurement time.  

\begin{figure}
\centering
\includegraphics[width=1.0\columnwidth,keepaspectratio]{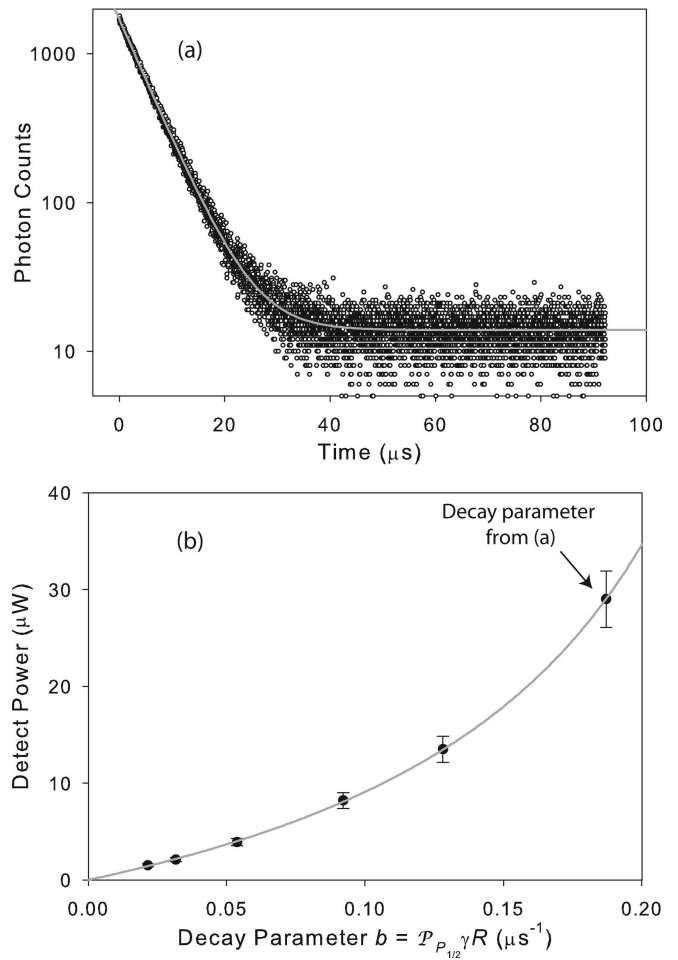}
\caption{Data used to determine the branching ratio of the ${}^2P_{1/2}$ level.  (a) Exponential decay in the fluorescence of a trapped $^{174}$Yb$^{+}$ ion, indicating population trapping in ${}^2D_{3/2}$.  During this time, $\approx$29 $\mu$W of 369.53 nm light was incident on the ion, with the 935.2 nm laser detuned approximately 3 GHz from the ${}^2D_{3/2} \leftrightarrow {}^3D[3/2]_{1/2}$ transition.  Data collection time was 5 minutes, and is analyzed with 16 ns binning.  The gray line is an exponential (plus background) fit to the data.  (b) Exponential fit decay parameters $b = \mathcal{P}_{P_{1/2}} \gamma R$ for a range of 369.53 nm laser intensities.  Error bars are taken to be $\pm$10\% the measured 369.53 nm power, to account for possible intensity drifts of the laser during the measurement.  The line is a fit to the function $p = (2 b p_{sat})/(\gamma R - 2 b)$ with two fit parameters: $p_{sat}$ and $\gamma R$.}
\label{fig:branching-ratio}
\end{figure}

Figure~\ref{fig:branching-ratio}(a) shows this exponential decay for 29 $\mu$W of incident 369.53 nm light, focused to a waist of $\approx$30 $\mu$m at the ion.  We repeat the measurement for a variety of 369.53 nm light intensities, with the resulting decay parameters $b = \mathcal{P}_{P_{1/2}} \gamma R$ shown in  Fig.~\ref{fig:branching-ratio}(b).  Given the expressions for $b$ and $\mathcal{P}_{P_{1/2}}$ above, the data is fit to $p = (2 b p_{sat})/(\gamma R - 2 b)$ with two fit parameters: $p_{sat}$ and $\gamma R$.  Using $\gamma = 1/(8.07 \pm 0.09 \mbox{ ns})$~\cite{pinnington:lifetime}, we determine the ${}^2P_{1/2}$ branching ratio to ${}^2D_{3/2}$ to be $R = 0.00501(15)$.  This number is consistent with prior theoretical calculations~\cite{migdalek:yb-core-pol} and experimental measurements~\cite{yu:lifetime}.  Knowledge of this branching ratio can be important in determining the theoretical limit on state detection of an Yb$^{+}$ qubit, as discussed in Sec.~\ref{sec:qubit}.

\section{\label{app:hyperfine}${}^{171}$Y\lowercase{b}$^+$ ${}^2D_{3/2}$ and ${}^3D[3/2]_{1/2}$ Hyperfine Splittings}

In determining the frequency needed to drive the ${}^2D_{3/2} \vert F=2 \rangle \leftrightarrow {}^3D[3/2]_{1/2} \vert F=1 \rangle$ transition (see Sec.~\ref{sec:qubit}), we measured the hyperfine splitting of the  ${}^2D_{3/2}$ and ${}^3D[3/2]_{1/2}$ levels of ${}^{171}$Yb$^+$.  To begin, the 935.2 nm laser is tuned far ($\approx$3 GHz) from the ${}^2D_{3/2} \vert F=1 \rangle \leftrightarrow {}^3D[3/2]_{1/2} \vert F=0 \rangle$ transition, stabilized to the reference cavity, and the power is reduced to approximately 20 $\mu$W at the trap.  We then scan the rf frequency applied to the fiber EOM in the 935.2 nm beam over a wide range (6 GHz).  During this frequency scan, the fluorescence of the trapped ${}^{171}$Yb$^{+}$ ion is monitored using the PMT.  As one of the first-order sidebands generated by the fiber EOM scans across either the ${}^2D_{3/2} \vert F=1 \rangle \leftrightarrow {}^3D[3/2]_{1/2} \vert F=0 \rangle$ or the ${}^2D_{3/2} \vert F=1 \rangle \leftrightarrow {}^3D[3/2]_{1/2} \vert F=1 \rangle$ resonance, many scattered 369.53 nm photons are observed.  The hyperfine splitting of the ${}^3D[3/2]_{1/2}$ level is the frequency difference between these two resonances, and is measured to be $2.2095(11)$ GHz.  

Next, we stabilize the laser at the ${}^2D_{3/2} \vert F=1 \rangle \leftrightarrow {}^3D[3/2]_{1/2} \vert F=0 \rangle$ transition.  While the ion is being cooled by the 369.53 nm light, we also apply a train of 2 ps pulses near 369.53 nm, generated by frequency doubling the pulses from a mode-locked Ti-sapphire laser~\cite{dave:lifetime,martin:ultrafast-rabi,peter:interference}.  These broadband ($\approx$250 GHz) pulses couple to both of the hyperfine levels in ${}^2S_{1/2}$.  Thus, the ion is continuously excited to the ${}^2P_{1/2} \vert F=1 \rangle$ manifold, which eventually decays to ${}^2D_{3/2} \vert F=2 \rangle$, resulting in population trapping in this level.  The first-order sidebands produced by the fiber EOM in the 935.2 nm beam are now scanned until one of the generated frequency components passes through resonance with the ${}^2D_{3/2} \vert F=2 \rangle \leftrightarrow {}^3D[3/2]_{1/2} \vert F=1 \rangle$ transition, signaled by an increased fluorescence of the ion.  Given the hyperfine splitting of the ${}^3D[3/2]_{1/2}$ level stated above, we determine the hyperfine splitting of the ${}^2D_{3/2}$ level to be $0.86(2)$ GHz.


\begin{thebibliography}{59}
\bibitem[{dou()}]{doubling-note}
\bibinfo{note}{As a result of the free spectral range of the doubling cavity,
  this modulation frequency does not appear in the 369.53 nm light.}
\bibitem[{sid()}]{sideband-note}
\bibinfo{note}{Since the fraction of power transferred into a first-order
  sideband is given by $(J_1(\theta))^2$, this is the maximum amount that can
  be expected. Here, $\theta$ is the phase-shift induced by the modulating
  voltage. The remaining light at other frequencies is far detuned from any
  iodine feature, and thus does not contribute to the absorption signal.}
\bibitem[{cra()}]{crap-eom-note}
\bibinfo{note}{Less than 1\% of the power is transferred into each of the
  second-order sidebands, limited by technical difficulties with the
  modulator.}
\bibitem[{pow()}]{power-note}
\bibinfo{note}{Far less power at 935.2 nm has been used with a negligible
  decrease in ion fluorescence.}
\bibitem[{pum()}]{pump-note}
\bibinfo{note}{Adequate rf power is applied to the EOM such that the carrier is
  almost completely suppressed.}
\bibitem[{det()}]{detect-note}
\bibinfo{note}{In Ref.~\cite{mark:detection}, a theoretical detection
  efficiency of 99.33\% is given. The discrepancy here arises from using
  different values for the lifetime of the ${}^2P_{1/2}$ excited state. In
  present calculation, we use 19.7 MHz, as found in
  Ref.~\cite{pinnington:lifetime}.}
\end{thebibliography}
\end{document}